\documentclass[fleqn,usenatbib]{mnras}

\usepackage[utf8]{inputenc}
\usepackage{enumitem}
\usepackage{graphicx}
\usepackage{hyperref}
\usepackage{mathtools}
\usepackage{comment}
\usepackage{amsmath}
\usepackage{ amssymb }
\usepackage{lipsum} 
\usepackage{float}
\usepackage{cleveref}
\usepackage[caption=false]{subfig}
\usepackage{placeins}
\usepackage{xcolor}

\def\s{{\rm s}}
\def\cm{{\rm cm}}
\def\kpc{{\rm kpc}}
\def\gev{{\rm GeV}}

\title[Gamma-ray ABC]{{Approximate Bayesian Computation Applied to the Diffuse Gamma-ray Sky}}

\author[E.~J. Baxter et al.]{
Eric J. Baxter,$^{1}$
J. G. Christy,$^{2}$
Jason Kumar$^{2}$
\\$^{1}$Institute for Astronomy, University of Hawai`i, 2680 Woodlawn Drive, Honolulu, HI 96822, USA
\\$^{2}$Department of Physics \& Astronomy, University of Hawai`i, 2505 Correa Road, Honolulu, HI 96822, USA
}

\begin{document}

\label{firstpage}

\pagerange{\pageref{firstpage}--\pageref{lastpage}}
\maketitle

\begin{abstract}
Many sources contribute to the diffuse gamma-ray background (DGRB), including star forming galaxies, active galactic nuclei, and cosmic ray interactions in the Milky Way.  Exotic sources, such as dark matter annihilation, may also make some contribution.  The photon counts-in-pixels distribution is a powerful tool for analyzing the DGRB and determining the relative contributions of different sources.  However, including photon energy information in a likelihood analysis of the counts-in-pixels distribution quickly becomes computationally intractable as the number of source types and energy bins increase.  Here, we apply the likelihood-free method of Approximate Bayesian Computation (ABC) to the problem.  We consider a mock analysis that includes contributions from dark matter annihilation in galactic subhalos as well as astrophysical backgrounds.  We show that our results using ABC are consistent with the exact likelihood when energy information is discarded, and that significantly tighter parameter constraints can be obtained with ABC when energy information is included.  ABC presents a powerful tool for analyzing the DGRB and understanding its varied origins.
\end{abstract}  


\section{Introduction}

The diffuse gamma-ray background (DGRB) is all gamma-ray photons on the sky that cannot be identified with individual sources above the detection threshold.  The DGRB receives contributions from point sources below the detection threshold, as well as truly diffuse emission from e.g. cosmic ray interactions in the Milky Way.  Several different classes of sources are known to contribute to the DGRB, including blazars \citep{Stecker:1993} and star forming galaxies \citep{Pavlidou:2002}.  More exotic sources, such as gamma rays produced in dark matter annihilation within the Milky Way and in extragalactic dark matter halos, may also contribute to the DGRB \citep[e.g][]{Fermi:2015}.  

Identifying the precise relative contributions of these and other sources to the DGRB is a long standing problem in gamma-ray astronomy \citep[for  a review, see][]{Fornasa:2015}.  Identifying the relative contributions of standard astrophysical sources to the DGRB would teach us about the properties of these sources.  A detection of exotic contributions to the DGRB, such as dark matter annihilation, would represent a major discovery of physics beyond the Standard Model.

Information about sources contributing to the DGRB is encoded in the spatial (i.e. directional) and energy distributions of the gamma-ray photons.  There are several ways that the spatial distribution of photons might be used to distinguish contributions to the DGRB.  For instance, a significant part of the anisotropic gamma-ray flux comes from regions near the galactic plane.  Distinguishing statistically isotropic sources --- including essentially all extragalactic sources --- is more challenging, but could be accomplished using statistics such as the power spectrum \citep[e.g][]{Harding:2012} and the cross-correlation with some external catalog \citep[e.g.][]{Xia:2011, Baxter:2022}.  The spectral or energy distribution of gamma-ray photons also contains information about different contributions to the DGRB.  Blazars and star forming galaxies, for instance, are known to have significantly different spectra \citep[e.g][]{Pavlidou:2002}, potentially enabling their contributions to be separated.  In the case of dark matter annihilation  the distribution of photon energies carries information about the dark matter particle properties \citep{Cirelli:2010xx}.

We focus here on separating the contributions to the DGRB from astrophysical backgrounds and from dark matter annihilation in subhalos within our galaxy.  Cold dark matter (CDM) models predict the existence of many small clumps of dark matter (subhalos) within the larger dark matter halo that hosts our galaxy \citep{Madau:2008}.  Since the dark matter annihilation rate scales with the square of the dark matter density, the enhanced dark matter density in subhalos makes them potentially promising targets for indirect detection searches. A detection of photons from annihilation in dark matter subhalos would constrain the properties of the dark matter and teach us about the small-scale structure of CDM.

A particularly powerful statistic for studying the DGRB is the photon counts in pixels distribution  ~\citep{Lee:2009,Baxter:2010,Malyshev:2011,Somalwar:2020awt,Runburg:2021}.
As a starting point, we could consider the histogram of photon counts in pixels across the sky, i.e. the number of pixels that contain $C$ photons, as a function of $C$.  This distribution encodes information about the spatial distribution of the photons.  Generalizing the idea of the photon counts histogram somewhat, we can then define the photon count probability distribution function (PDF) as the probability, $P_{i,C}(C)$, of observing $C$ photons in the $i$th pixel.  The fact that the photon count PDF is a powerful tool for analyzing the DGRB can be seen with the following example.  Consider two source classes, one that is rare and bright, and another that is numerous and faint.  Even if these two source classes were to yield the same mean flux on the sky or the same angular power spectrum, they would have very different photon count PDFs.   The faint and numerous sources are expected to generate a Poisson-like $P_{i,C}(C)$, while the $P_{i,C}(C)$ for rare and bright sources is expected to be significantly non-Poisson.  As we discuss below, the photon count PDF is directly sensitive to the population statistics of a source class.  The gamma-ray photon count PDF from dark matter annihilation in subhalos has been considered by several authors \citep{Lee:2009,Baxter:2010,Somalwar:2020awt,Runburg:2021}.

We note that the counts PDF --- also called the points-in-cells or counts-in-pixels distribution --- has a long history of application to the study of unresolved source populations outside of gamma-ray astronomy.    In radio astronomy, the technique is sometimes called $P(D)$ analysis \citep[e.g][]{Scheuer:1957}, and in X-ray astronomy, it often goes by the name of fluctuation analysis \citep{Hasinger:1993}.

The standard Bayesian approach to analyzing the photon count distribution involves first computing the likelihood, $\mathcal{L}(d | \theta)$, which represents the probability of observing the data, $d$, given model parameters, $\theta$.  In the case of the photon count distribution, the data can be in the form of a photon counts map (i.e. the number of photons observed in pixels across the sky), or compressed further into a summary statistic like the histogram of photon counts.  We describe the computation of such a likelihood in \S\ref{sec:with_likelihood}.  Given a likelihood and some prior information on the parameters represented by the probability distribution, $\mathrm{Pr}(\theta)$, parameter constraints can be obtained in Bayesian fashion by sampling from the posterior, $P(\theta | d) \propto \mathcal{L}(d | \theta)\mathrm{Pr}(\theta)$ using techniques such as Markov Chain Monte Carlo (MCMC).  Given these samples, credible intervals on the parameters of interest can be easily obtained.

However, as we describe below, computing the exact likelihood for the photon counts distribution becomes increasingly difficult, if not intractable, when there are more than one source class and more than one photon energy bin.  This is unfortunate, since energy information has the potential to add significant discriminatory power, and gamma-ray detectors, such as the Fermi Gamma-Ray Space Telescope,\footnote{\url{https://fermi.gsfc.nasa.gov/}} are typically capable of measuring both photon directions and energies.  In order to feasibly include energy information in an analysis, it is therefore useful to consider so-called {\it likelihood-free} methods that do not require calculation of an explicit likelihood.\footnote{Since these method often involve approximation of the likelihood, they are perhaps more aptly termed {\it simulation-based} rather than likelihood-free \citep{Cranmer:2020}.  }  Here, we consider the application of the likelihood-free method of approximate Bayesian computation (ABC) to separate different contributions to the DGRB. 
To our knowledge, this is the first application of ABC to the problem.

ABC is useful in situations in which it is  
extremely difficult to compute the exact likelihood, but it 
is fairly straightforward to 
simply generate a mock data set.  This can occur, for 
example, in any situation in which many different 
processes can produce the same data.  Determining the likelihood of any 
particular data requires summing probabilities over all 
of the many processes which could produce that data, 
which in turn involves sums over combinatoric possibilities which are 
computationally cumbersome.  
Generating mock 
data however is simply a matter of choosing a process 
based on a probability distribution, and then choosing 
the output of this process from another probability 
distribution; the combinatoric sums are absent.  

In its simplest incarnation, the idea of ABC is that one simply scans over the parameter 
space, generating mock data sets, and measuring how 
close those mock data sets are to actual data using a 
distance metric.  In a nutshell, by creating many 
mock data sets and locating those points in parameter 
space for which the mock data is sufficiently 
similar to the actual data, one has essentially 
found the region of parameter space with 
high likelihood of yielding the data, even though the 
likelihood function itself remains unknown.  This algorithm is known as rejection sampling.  Below, we will explore more sophisticated sampling algorithms, but the basic idea of generating simulated data and computing a distance metric with respect to the true data will remain the same.

In our particular application of ABC, we consider two types of photon sources --- dark matter subhalos and astrophysical backgrounds --- each of which emits photons with a different energy spectrum.  In this case, computing an  exact likelihood is intractable when the number of energy bins is large, since the probability  of a photon having any particular energy depends on which source produced it.  Thus, computing the likelihood  of observing a large number of photons with particular energies requires a sum over all ways of partitioning those photons among the sources and energy bins (\S\ref{sec:exact_likelihood}).  On the other  hand, it is fairly straightforward to generate a mock data set  by simply drawing the number of photons emitted by each source in a pixel from a probability distribution, and then drawing the energies of each photon using the appropriate energy spectrum for each source.

One could imagine using other likelihood-free methods to infer the composition of the gamma-ray sky besides ABC.  Recently, \citet{Michra-Sharma:2021} presented an application of the likelihood-free method of density estimation using normalizing flows to an analysis of the galactic center gamma-ray excess in order to determine how much of this excess could be attributed to unresolved point sources.  In the normalizing flows approach, machine learning methods --- typically deep neural networks --- are used to directly model (i.e. learn) the posterior distribution on the parameters by training with a set of simulated data realizations generated at parameter values drawn from some prior distribution \citep{Rezende:2015,Papamakarios:2019}.  \citet{Michra-Sharma:2021} additionally use a neural network-based approach to learn optimized summary statistics from the data.  In addition to the methodological differences between our work and that of \citet{Michra-Sharma:2021}, we note that our focus is on the inclusion of energy information into an analysis of the photon counts PDF.  \citet{Michra-Sharma:2021}, on the other hand, do not include photon energy information in their analysis.

The plan of this paper is as follows.  In \S\ref{sec:with_likelihood}, we describe the exact likelihood for the energy-dependent photon counts distribution, showing that it quickly becomes intractable for multiple sources and multiple energy bins.  In \S\ref{sec:likelihood_free}, we discuss how ABC can be applied to this problem, making it possible to generate parameter constraints even in multi-energy bin, multi-source scenarios.  In \S\ref{sec:PDF_models}, we describe our models for the photon count distributions from two sources: dark matter annihilation in galactic subhalos and astrophysical backgrounds.  In \S\ref{sec:spectral_models} we describe our spectral models for these sources.  We demonstrate the application of ABC to the problem in \S\ref{sec:application}, showing that it recovers the exact posterior in the case of a single energy bin and multiple sources, and that significantly tighter parameter constraints can be obtained when energy information is included.  We conclude with a discussion of our results in \S\ref{sec:discussion}.

\section{Analyzing photon count data with a likelihood}
\label{sec:with_likelihood}

We consider photon data that has been binned into spatial pixels and energy bins.  In other words, we measure a set of integers, $d_{i\alpha}$, representing the number of photons that were observed in the $i$th pixel in a map, and that had energies, $E$, in the range $E_{\alpha}^{\rm low} < E < E_{\alpha}^{\rm high}$, where $E_{\alpha}^{\rm low}$ and $E_{\alpha}^{\rm high}$ are the limits of the $\alpha$th energy bin.  Pixelizing the data in this way should not result in significant information loss, provided the size of the pixels is well matched to the beam of the telescope, and the size of the energy bins is well matched to the energy resolution of the detector.

To model the data, we assume that for every source type of interest (e.g. galaxies, dark matter annihilation, etc.) one can compute the photon count PDF and the spectrum (we will discuss the calculation of the PDF in \S\ref{sec:PDF_models}).  We represent the photon count PDF with $P^s_{C,i}(C)$, which describes the probability of observing $C$ photons from source type $s$ in the $i$th pixel.  Note that these photons can originate from multiple objects of the same source type within a pixel.  We represent the spectrum with $f^s_{E,i}(E)$, which describes the relative probability of observing photons of energies $E$ in pixel $i$ from source $s$.  Note that for statistically isotropic sources, there is no dependence of $P^s_{C,i}$ or $f^s_{E,i}$ on $i$. 

For simplicity, in the mock analyses presented here, we will assume that the photon counts in each pixel are statistically independent.  In other words, every draw from $P^s_{C,i}(C)$ is independent of every other draw.  This assumption could be violated by, for instance, an instrumental beam which redistributes photons from one direction into several pixels. However, it is an acceptable approximation, since pixel-to-pixel correlation can always be reduced by adopting larger pixel sizes.  We plan to revisit this assumption in future work.

We now consider how the data can be analyzed to extract constraints on the parameters of these models.  We begin with the computation of the exact likelihood.

\subsection{Exact likelihood: no energy information}
\label{sec:exact_likelihood}

We first consider the case where photon energy information is not used in the analysis, which is equivalent to having only a single energy bin. We denote the number of source classes as $N_S$.  The total counts PDF for the $i$th pixel, $P^{\rm tot}_{C,i}(C)$, can be expressed as a convolution over the PDFs for the sources contributing to that pixel:
\begin{multline}
\label{eq:pc_tot}
P^{\rm tot}_{C,i}(C) = \sum_{k_1 = 0}^C \sum_{k_2 = 0}^C \cdots \sum_{k_{N_S} = 0}^C 
\left[ \prod_{s=1}^{N_S} P^{s}_{C,i}(k_s)  \right]
\\
 \times \delta \left(C - \sum_{s=1}^{N_S} 
k_s \right),
\end{multline}
where we have used $s$ to index the different source types,  $k_s$ represents the counts from source $s$, and the $\delta$ function enforces the fact that the sum of photons from all sources should be $C$.  

Assuming that the photon counts in each pixel are independent, the total likelihood for the pixelized map, $\vec{d}$, is simply
\begin{equation}
\label{eq:one_bin_likelihood}
    \mathcal{L}(\vec{d} | \vec{\theta}) =  \prod_i^{N_{\rm pix}} P^{\rm tot}_{C,i}(d_i | \vec{\theta}),
\end{equation}
where the product runs over all $N_{\rm pix}$ pixels in the map and $d_i$ is the total photon counts in the $i$th pixel,
\begin{equation}
d_i = \sum_{\alpha} d_{i\alpha}.
\end{equation}

\subsection{Exact likelihood: with energy information}
\label{sec:exact_likelihood_wenergy}

The likelihood becomes significantly more complicated upon including energy information.  To illustrate this, it is sufficient to consider a single pixel and two source 
classes, signal ($A$) and background ($B$).  Suppose there are $N_E$ energy bins, indexed by 
$\alpha$, and the probability of a photon produced by source $s$ having an energy 
lying in bin $\alpha$ 
is $f^{s}_{\alpha}$, where $\sum_{\alpha = 1}^{N_E} f^{s}_{\alpha} =1$.
Then the probability of source $A$ producing a photon count $c_\alpha$ in 
each energy bin $\alpha$ is given by
\begin{multline}
P^A( \{c_1 ,...,c_n \} ) = P^A_{C} \left(\sum_{\alpha =1}^{N_E} c_\alpha \right) 
\\
 \times
\left( \prod_{\alpha=1}^{N_E} (f^{A}_{\alpha})^{c_\alpha} \right) 
\frac{\left( \sum_{\alpha =1}^{N_E} c_\alpha \right)!}{\prod_{\alpha=1}^{N_E} c_\alpha!}.
\end{multline}
The likelihood for the observed energy-dependent counts in the pixel, $\vec{d}_i \equiv \{d_{i0}, d_{i1},\ldots,d_{i N_E}\}$, is then given by
\begin{multline}
\label{eq:e_counts_single_source}
\mathcal{L}(\vec{d}_i|\vec{\theta}) = \\
\sum_{c_1 =0}^{d_{i1}} ... \sum_{c_{N_E} =0}^{d_{iN_E}} 
P^A ( \{c_1, ..., c_{N_E} \}) P^B (\{d_{i1} - c_1, ..., d_{iN_E} - c_{N_E}  \}) .
\end{multline}

In order to fully utilize the energy information, one would need the number of energy bins ($N_E$) to be large, leading to a large number of convolutions in the expression for the likelihood.  If all photon count distributions $P^s_{C}$ are Poisson, then the convolutions in Eq.~\ref{eq:e_counts_single_source} become trivial, since the convolution of two Poisson distributions is also a Poisson distribution.  But 
if one or more  sources has a non-Poisson photon count distribution, then the convolutions must be performed, quickly leading to calculations which are computationally intractable.  This problem becomes even more severe if we have more than two source classes. 

\section{Analyzing the photon count distribution without a likelihood}
\label{sec:likelihood_free}

The above discussion demonstrates that computing the exact likelihood for the energy-dependent photon counts may be computationally intractable when there are as few as two sources.  In essence, this difficulty emerges from the fact that the same observed photon count distribution in energy bins could be achieved in many different ways from a combination of two sources.  This means that going from the observed counts to the likelihood requires summing over many combinatoric possibilities.  On the other hand, generating a mock realization of the data given a model for the counts distributions, $P^{s}_C(C)$, and spectra, $f^s_E(E)$, for several sources is trivial.  One simply draws from the $P^{s}_C(C)$ for each source, randomly assigns photons energies from the $f^s_E(E)$ for each source, and then sums the resulting counts from each source to produce the final map.  

A situation in which computing the exact likelihood is computationally intensive,  but simulating the data is trivial presents an ideal application for approximate Bayesian computation \citep[ABC;][]{Rubin:1984}.  In ABC, one generates mock realizations of the data at a point in parameter space, and compares these realizations (optionally compressed to a summary statistic) to the true data using some distance metric.  Points in parameter space that yield mock data close to the true data are then up-weighted relative to points that do not, resulting in an estimate of the posterior.  At no point is a likelihood computed.

There are three ingredients to an implementation of ABC: (1) a choice of summary statistic, (2) a choice of distance metric, and (3) a means for exploring the parameter space and keeping/rejecting parameter samples.  We now describe in more precise terms our implementation of ABC with regard to these three ingredients, and the application of ABC to the problem at hand.

\subsection{Choice of summary statistic}
\label{sec:summary_stat}

The data that we consider in this analysis are pixelized maps of the sky in several different energy bins, $d_{i\alpha}$, where $i$ indexes the pixel and $\alpha$ indexes the energy bin.  In principle, the comparison between the observed data and the mock data generated during the ABC process could occur in the map-space.  However, for the present analysis, we expect to have thousands of pixels and $N_E = 10$ energy bins, making the dimensionality of the data very high.  The so-called curse of dimensionality means that sampling in such a high-dimensional space will be very inefficient, with vanishingly few samples passing any distance threshold.  It is therefore desirable to compress the data into some lower dimensional summary statistic which preserves most of the information in the original data.

We will use the energy-dependent histogram of photon counts as the summary statistic in our analysis.  In other words, for each energy bin $\alpha$, we count the number of pixels, $h_{\alpha\gamma}$, that have photon counts in some count bin $C^{\rm low}_{\alpha\gamma} < C < C_{\alpha\gamma}^{\rm high}$, where $C_{\alpha\gamma}^{\rm low}$ and $C_{\alpha\gamma}^{\rm high}$ are the bin edges and $\gamma$ indexes the count bin.  We keep the number of counts bins, $N_C$, fixed for every energy bin.  The dimension of the summary statistic is therefore $N_E \times N_C$.  However, since the spectra of the signal and backgrounds decrease rapidly with increasing photon energy (see discussion of the spectral models in \S\ref{sec:spectral_models}), the number of photons produced at high energies is much smaller than at low energies.  Consequently, we decrease the maximum count value of the histogram with increasing energy.  We present the summary statistic computed for an example map in \S\ref{sec:application}.

The energy-dependent histogram of photon counts effectively throws away all spatial information in the data.  If there are large spatial variations, this approach would clearly be suboptimal, and other summary statistics might be warranted.  However, as we describe below, the sources that we consider in our analysis are statistically isotropic.  Therefore, we do not expect to lose a significant amount of information by using the histogram.  Some information may be lost by using a small number of count bins or energy bins.  Additionally, the histogram does not optimally compress the data, since some histogram bins may only be populated for very extreme values of the parameters.    More optimal summary statistics could be identified using machine learning methods, as discussed in \citet{Michra-Sharma:2021}.  However, as we show below, the energy-dependent histogram appears sufficient to obtain tight parameter constraints in a reasonable runtime.  The histogram has the added advantage of easy interpretation.

\subsection{Choice of distance metric}

Given a mock data realization corresponding to a point in parameter space, the determination of whether to keep or reject that parameter point is made on the basis of the distance between the mock data and the true data.  This distance is evaluated in the space of the summary statistic, which in our case is the (energy-dependent) histogram.  Many choices of distance metric are possible.  Common choices include, for example, $L_1$ and $L_2$ norms.  

In the case of the histogram summary statistic, fluctuations about the mean histogram are expected to be approximately Poisson distributed, even if the photon counts distribution, $P_C(C)$, is non-Poisson.  This can be seen with the following argument.  Consider the probability of there being $h$ pixels with photons counts $\gamma$. Assuming, as we have, that the photon count distribution is isotropic,  this probability is given by
\begin{eqnarray}
    P(h) &=& P_C(\gamma)^h (1 - P_C(\gamma))^{N_{\rm pix} - h} \frac{N_{\rm pix}!}{h!(N_{\rm pix} - h)!} \\
    &=& \left( \frac{P_C(\gamma)}{1-P_C(\gamma)} \right)^h \frac{N_{\rm pix}!}{h!(N_{\rm pix} - h)!} (1 - P_C(\gamma))^{N_{\rm pix} } \nonumber  \\
    &\approx& (P_C(\gamma))^{h} \frac{N_{\rm pix}^{h}}{h!} \exp(-P_C(\gamma)N_{\rm pix}) \\
    &=& \mathcal{P}(h | N_{\rm pix}P_C(\gamma)),
\end{eqnarray}
where we have assumed $P_C(\gamma) \ll 1$ and $N_{\rm pix} \gg h$, and $\mathcal{P}$ is the Poisson distribution.  Thus, fluctuations in $h_{\alpha\gamma}$ are expected to be approximately Poisson distributed.

Since the variance of a Poisson distribution is the same as its mean, if $h_{\alpha\gamma}$ is large, then its variance will be large as well.  A distance metric which takes this into account is the $\chi^2$ distance \citep[e.g.][]{Pele:2010}.  For two histograms $h$ and $h'$ this distance is given by
\begin{equation}
    d(h_1, h_2) = \sqrt{\sum_{\alpha,\gamma} \frac{(h_{\alpha\gamma} - h'_{\alpha\gamma})^2}{h_{\alpha\gamma} + h'_{\alpha\gamma}}}.
\end{equation}
The term in the denominator is effectively the expected variance of the difference between the two histograms.  We use this distance measure in our implementation of ABC. 

\subsection{Parameter sampling}

The final ingredient needed for ABC is some method of sampling the parameter space.  The most basic approach to this is  rejection sampling.  One begins by drawing many parameter points from the 
prior distributions.  For each of these 
parameter points, one generates a mock data set.
If the distance $d$ between a mock data set and the 
true data set is less than a user-chosen threshold 
$\epsilon$, then the corresponding parameter point is 
kept, while the remaining parameter points are discarded.  In the limit that $\epsilon \rightarrow 0$, 
the normalized density of parameter points which survive rejection sampling will converge to the exact posterior, but as $\epsilon$ becomes smaller, it will take longer to perform the sampling (since it is unlikely for a mock data set to agree exactly with the true data by chance).

A more sophisticated approach that improves sampling efficiency is Sequential Monte Carlo ABC \citep{Sisson:2007}.  In this case, the proposal distribution used to draw parameter points is updated during the course of sampling.  Initially, the proposal distribution may be set to the prior, and one can use a large value of the distance threshold, $\epsilon$.  After some number of samples are accepted, the proposal distribution is updated to reflect the new information about the posterior, $\epsilon$ is reduced, and the process is repeated. 
Typically the set of proposal points are referred to as \textit{particles}.  In the ABC population Monte Carlo (PMC) approach suggested in \citet{Beaumont:2008}, the particles are updated by translating the particles from a previous iteration using some kernel function.  These particles are then assigned importance weights based on the previous distribution of particles.

Here we rely on a modified version of the ABC PMC algorithm based on \citet{Simola:2019} and implemented in the code \texttt{ELFI} \citep{ELFI}.  The \citet{Simola:2019} approach provides a rule for selecting the series of tolerances $\epsilon_1>...>\epsilon_n>0$ used in each iteration.  These tolerances are chosen on the basis of quantiles, i.e. the tolerance is set so that some chosen quantile of points have distance less than threshold.  The \citet{Simola:2019} method adjusts the quantiles based on how much the posterior changes from one iteration to the next.  If the posterior has changed significantly, the quantile can be reduced more.  If the posterior has not changed at all, the quantile will approach 1.

\section{Modeling the photon count distribution}
\label{sec:PDF_models}

To illustrate the application of ABC to the DGRB we will consider a mock analysis that includes contributions to the sky from astrophysical backgrounds and from dark matter annihilation in galactic subhalos.  Such an analysis requires developing models for the $P_C(C)$ from these two source classes, which we now describe.  While we focus on the case of dark matter annihilation in galactic subhalos, the techniques described here could be easily extended to other source populations.

\subsection{From source statistics to photon count distributions}
\label{sec:pdf_calculation}

We first describe in a general fashion how to go from a model for the population statistics of a source to the photon count distribution, $P_C(C)$.  Later, we will specialize to the case of dark matter annihilation in subhalos.  More detailed expositions of these methods can be found in \citet{Lee:2009} and \citet{Baxter:2010}.

We begin by computing $P_1(F)$, the probability of observing a flux $F$ from a single source along a line of sight.  This probability is given by
\begin{equation}
\label{eq:p1f}
    P_1(F) \propto \int d\ell~ \ell^2 \int dL \int dx \frac{d^2N}{dx dV} P_L(L | x) \delta \left(F - \frac{L}{4\pi \ell^2} \right),
\end{equation}
where $\ell$ is the line of sight distance and $L$ is the luminosity of a source.  We use $x$ to represent the quantity (such as halo mass) that is used to express the differential abundance, $d^2N/dxdV$.  The probability distribution $P_L(L|x)$ then represents the probabilistic relationship between $L$ and $x$.

The average number of sources in a solid angle $\Omega$ is then given by
\begin{equation}
    \mu = \Omega \int d\ell ~ \ell^2 \int dx \frac{d^2N}{dx dV}.
\end{equation}
We assume that the actual number of sources within $\Omega$, $n$, is a Poisson random variable  with expectation $\mu$.  The probability of obtaining a total flux $F$ in some direction, $P(F)$, is then given by $n$ convolutions of $P_1(F)$, weighted by the probability of having $n$ sources within 
$\Omega$, summed over all $n$.  As shown in \citet{Lee:2009}, $P(F)$ can be written as
\begin{equation}
    P(F) = \mathcal{F}^{-1} \{ e^{\mu \left(\mathcal{F} \{ P_1(F)\}  - 1 \right) }\},
\end{equation}
where $\mathcal{F}$ is the Fourier transform and $\mathcal{F}^{-1}$ is the  inverse Fourier transform, appropriately normalized.

The photon count distribution $P_C (C)$ is then the convolution of the total flux distribution 
$P(F)$ with the probability of observing $C$ photons given a flux $F$.  We will assume that 
this latter probability distribution is Poisson, with a mean number of counts given by 
$F \times ({\rm exposure})$.  We thus find
\begin{equation}
P(C) = \int dF~ P(F) \times {\cal P}(C|F \times ({\rm exposure})),
\end{equation}
where $\mathcal{P}(C|\bar{x})$ is the Poisson probability of $C$ counts when the expectation value is $\bar{x}$.

\begin{figure*}
    \centering
    \includegraphics[scale=0.48]{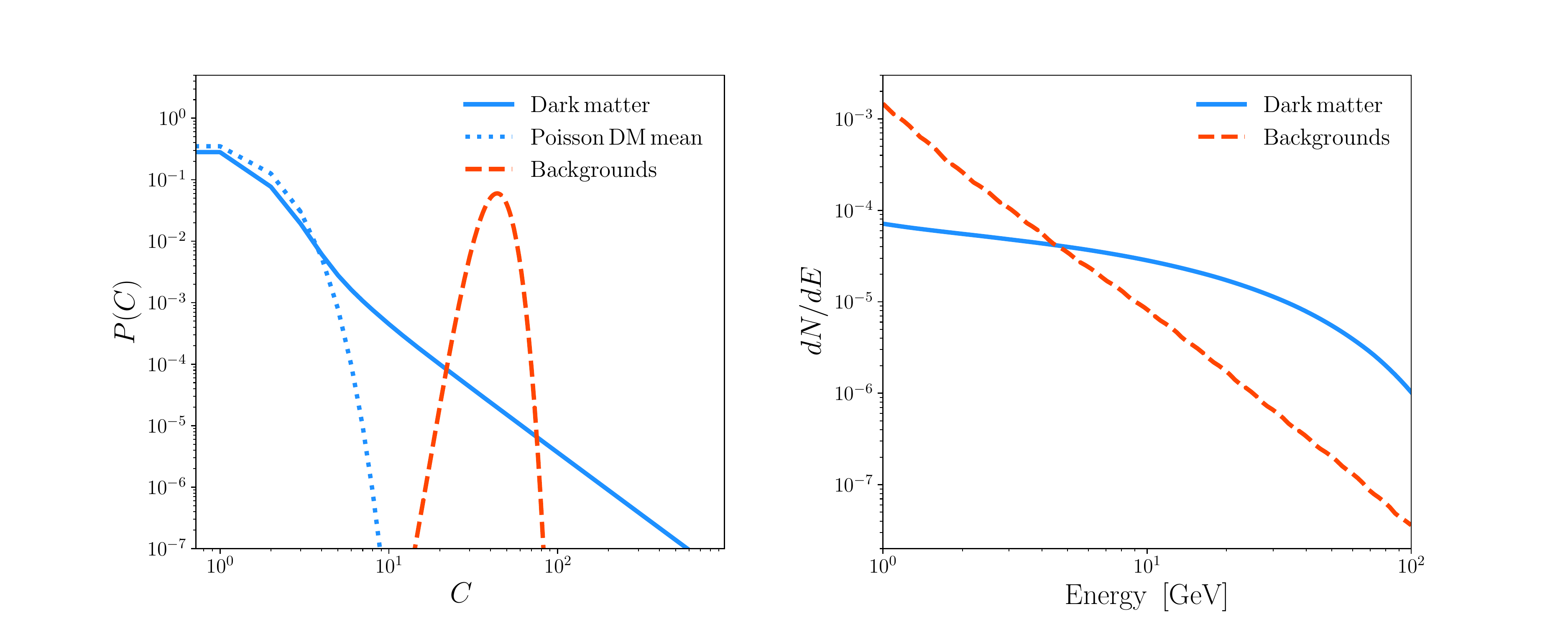}
    \caption{Illustration of the PDF (left) and spectral models (right) discussed in \S\ref{sec:PDF_models} and \S\ref{sec:spectral_models}, respectively.  Our baseline dark matter model (blue solid) assumes a $m_{\chi}$ particle that annihilates to $\tau\bar\tau$ in subhalos with minimum mass of $1 M_{\odot}$.  We consider an alternate dark matter model in  Appendix~\S\ref{app:bbar}.  Our background model (red dashed) is derived from estimates produced by the Fermi collaboration, as discussed in \S\ref{sec:astro_counts} and \S\ref{sec:astro_spectra}.  We assume an observation time of 10 years, a Fermi-like detector, and pixels that are roughly $1^{\circ}$ in size.  The blue dotted curve corresponds to a Poisson counts distribution model with the same mean as the dark matter signal; the true dark matter signal (blue solid) exhibits a non-Poisson power-law tail}.
    \label{fig:pdf_spectrum_models}
\end{figure*}

\subsection{Photon counts from dark matter subhalos}

We now apply the methods developed above to the case of dark matter annihilation in galactic subhalos.  Our treatment here follows that described in \citet{Lee:2009,Baxter:2010, Runburg:2021}.  We express the abundance of dark matter subhalos as a function of their mass:
\begin{equation}
\label{eq:mass_func}
\frac{d^2 N(r)}{dM dV} = A \frac{(M/M_\odot)^{-\beta}}
{\tilde r (1+\tilde r)^2} ,
\end{equation}
where $\tilde r \equiv r/r_s = r / 21~\kpc$ and $r_s$ is the scale radius of the Milky Way,
$A = 1.2 \times 10^4 \, M_\odot^{-1} \kpc^{-3}$, 
and $\beta = 1.9$.   We further assume that the mass function goes to zero for $M < M_{\rm min}$, where $M_{\rm min}$ is the minimum subhalo mass. 
The minimum subhalo mass is unknown, but is predicted to be as low as $10^{-9}\,M_{\odot}$ in some cold dark matter models \citep[e.g.][]{Martinez:2009}.  However, such microhalos will not necessarily survive to the present day, particularly within the Milky Way environment \citep{Zhao:2007}. 
Here we will set $M_{\rm min} = 1\,M_{\odot}$.

Following \citet{Koushiappas:2010}, we relate the annihilation luminosity of a subhalo, $L_{\rm sh}$, to the subhalo mass via a lognormal probability distribution
\begin{equation}
    P_L \left( \ln L_{\rm sh} | M, r (\ell, \psi_i)\right) =
\frac{1}{\sqrt{2\pi} \sigma} \exp \left[- 
\frac{(\ln L_{\rm sh} - \langle \ln L_{\rm sh} \rangle)^2}{2\sigma^2} 
\right],
\end{equation}
where the expectation value of the subhalo log-luminosity is given by
\begin{multline}
\langle \ln \left( L_{\rm sh}
/ \s^{-1} \right) \rangle =
77.4 + 
 0.87 \ln (M / 10^5 M_\odot)  \\
-0.23 \ln (r / 50\kpc) 
+ \ln \left(\frac{8\pi \Phi_{PP}}{10^{-28} \cm^3 \s^{-1} 
\gev^{-2}} \right),
\end{multline}
and
\begin{equation}
\sigma = 0.74 - 0.003\ln(M/10^{5}\,M_{\odot}) + 0.011 \ln(r/50\,{\rm kpc}).
\end{equation}
The parameter $\Phi_{PP}$ controls the amplitude of the annihilation signal, and is set by the particle physics properties of the dark matter via:
\begin{equation}
\Phi_{PP} = \frac{A_{\rm DM} \langle \sigma v\rangle_0}{8\pi m_{\chi}^2} \int_{E_{\rm min}}^{E_{\rm max}} dE_{\gamma} \frac{dN_{\gamma}}{dE_{\gamma}},
\end{equation}
where $dN/dE_{\gamma}$ the spectrum of photons produced per annihilation of the dark matter as a function of photon energy $E_{\gamma}$, $E_{\rm min}$ and $E_{\rm max}$ are the energy limits of the detector, $m_{\chi}$ is the mass of the dark matter particle.  We set $\langle \sigma v\rangle_0 = 3\times 10^{-26} \, {\rm cm}^{3}{\rm s}^{-1}$ (i.e. the cannonical WIMP cross-section).  The free parameter $A_{\rm DM}$ then sets the normalization of the dark matter signal.  We will discuss specific models for $dN/dE_{\gamma}$ in \S\ref{sec:spectral_models}.

While the model for the subhalo distribution in Eq.~\ref{eq:mass_func} is isotropic around the galactic center, the Sun is not at the galactic center.  Therefore, the distribution of annihilation radiation from galactic subhalos will not be isotropic.  However, since the scale radius of the Milky Way is significantly larger than the distance of the Sun to the galactic center, assuming isotropy is not a bad approximation.\footnote{Note that the distribution of annihilation radiation from subhalos will generally be more isotropic than that from the main halo.  For subhalos, the emission is proportional to $d^2N /dMdV \propto \left[\tilde{r}(1+\tilde{r})^2 \right]^{-1}$, while for the parent halos, the emission is proportional to the density squared, or $\left[\tilde{r}(1+\tilde{r})^2 \right]^{-2}$, for an Navarro-Frenk-White profile.}  For simplicity, we treat the subhalo annihilation radiation as isotropic, by assuming that the subhalo mass function within each pixel is the same.  In particualr, the photon count PDF is calculated assuming that the distance from the Sun to the galactic center is 8.5~kpc, and that the angle between all pixels and the galactic center is $40^{\circ}$.

The photon counts PDF that results from this model is shown with the solid blue curve in the left panel of Fig.~\ref{fig:pdf_spectrum_models}.   To illustrate that this PDF departs significantly from a Poisson PDF, we also show a Poisson distribution with the same mean (dashed blue curve).

\subsection{Photon counts from standard astrophysical sources}
\label{sec:astro_counts}

There are many known non-dark matter contributors to the gamma-ray sky, including gamma rays produced in cosmic ray interactions in the Milky Way, active galactic nuclei in other galaxies, and star forming galaxies.  We include a rough prescription for these sources in our analysis, as described below.  Since the main intent of this work is to demonstrate the utility of ABC to the analysis of the diffuse gamma-ray background, more realistic modeling of these backgrounds is not essential.  Including more accurate background prescriptions would change our quantitative results, but would not qualitatively change our methods.

The dominant source of diffuse gamma-rays within the galaxy is the interaction of high-energy cosmic rays with galactic matter and radiation.  Along a given line of sight, there are many parcels of gas and radiation that each have some small probability of producing a gamma ray through a cosmic ray interaction.  Consequently, along each line of sight, the statistics of the galactic gamma-ray emission will be described by a Poisson distribution.  In the language of the discussion above, provided the variance of $P_1(F)$ is finite, as one takes the limit $\mu \rightarrow \infty$, keeping $\mu \langle F_1 \rangle = 
\mu \int dF~F \times P_1 (F)$ fixed, then $P(F) \rightarrow \delta (F - \mu \langle F_1 \rangle)$, 
and $P(C)$ becomes Poisson. 

Extragalactic contributions to the gamma-ray sky are dominated by emission from star forming galaxies and from blazars.  Since the number of star forming galaxies in a given pixel is likely to be large, the photon count distribution from these sources is expected to be close to Poisson \cite{Malyshev:2011}. For other source classes, blazars in particular, the photon count distribution is not expected to be Poisson \citep[e.g.][]{Malyshev:2011}.  For simplicity, we ignore the non-Poisson nature of these sources here.  This is likely an acceptable approximation, since blazars are expected to make a subdominant contribution to the DGRB compared to star forming galaxies \citep{Malyshev:2011, Roth:2021}.

We will also significantly simplify our analysis by assuming that the total background contribution is isotropic.  This is a reasonable approximation far from the galactic plane and from the galactic center.  We will therefore restrict our analysis to the parts of the sky that are more than $30^{\circ}$ from the galactic plane, and more than $60^{\circ}$ from the galactic center.  The resultant mask can be seen in Fig.~\ref{fig:example_map}.

Including non-isotropic background models would not substantially change our methodology.  In particular, ABC would still provide a solution to the problem of analyzing the energy-dependent photon counts distribution.  Anisotropy could, however, make the use of the energy-dependent histogram summary statistic suboptimal,  since the histogram effectively erases directional information.  As noted previously, though, our main intent is to demonstrate that likelihood-free inference provides a solution to the analysis of the energy-dependent photon counts distribution.  We therefore postpone consideration of anisotropic sources and alternative summary statistics to future work.

Our estimate for the total non-dark matter gamma-ray backgrounds is derived from the energy-dependent 
isotropic and galactic background models 
produced by the Fermi collaboration.\footnote{\url{https://fermi.gsfc.nasa.gov/ssc/data/access/lat/BackgroundModels.html}}   The galactic model is constructed by fitting several templates for the emission to the observed gamma-ray sky.  These templates are derived from ancillary observations in combination with models for cosmic ray propagation throughout the galaxy. The isotropic model is effectively the isotropic residual that results from the template fitting process.   The isotropic model will include unresolved extragalactic gamma ray and the residual charged particle background (i.e. events misclassified as gamma rays).  We sum the isotropic and galactic background models, and then average the resulting model across the unmasked region of the sky to arrive at an effective isotropic background flux.  Note that this procedure removes spatial variation from the background model, but preserves the total photon statistics.  We expect the latter to be most relevant for determining the accuracy of our forecasts.

As described above, we assume that this background is Poisson distributed.  The photon counts PDF that results from the galactic model is shown as the red dashed curve in the left panel of Fig.~\ref{fig:pdf_spectrum_models}.

\section{Spectral models}
\label{sec:spectral_models}

To incorporate energy information into our analysis, we must model the spectra of the dark matter and background contributions to the DGRB.  We restrict our analysis to photon energies in the range 1~GeV to 100~GeV, roughly corresponding to the sensitivity of Fermi, in the 
regime in which its effective area is independent of energy.

\subsection{Dark matter annihilation spectral models}
\label{sec:darkmatter_spectra}

As a typical benchmark scenario, we consider the well-motivated case in which dark matter annihilates 
to $\bar \tau \tau$ with $\sim 100\%$ branching fraction.  This scenario can arise, for example, if dark 
matter is its own anti-particle, and couples most strongly to Standard Model leptons through interactions 
which respect Minimal Flavor Violation (see, for example~\citealt{Kumar:2013iva}).  In this case, the 
$s$-wave annihilation cross section to a fermion/anti-fermion pair ($\chi \chi \rightarrow \bar f f$) 
is chirality-suppressed by a factor $(m_f / m_\chi)^2$, and one thus expects the branching fraction to be 
largest to the heaviest kinematically-accessible fermion.  If dark matter couples most strongly to leptons, 
then one expects the branching fraction 
$\bar \tau \tau$ to be $\sim 100\%$.

The spectrum of 
photons produced per annihilation, as a result of 
final state radiation and hadronic cascade decays, 
can be determined from {\tt PPPC4DMID}~\citep{Cirelli:2010xx} as a function of the 
dark matter mass.  
In the right panel of Figure~\ref{fig:pdf_spectrum_models}, we plot the photon spectrum for the $\bar \tau \tau$ channel assuming 
$m_\chi = 200~\gev$ (blue).  The photon spectrum of the $\bar \tau \tau$ channel with $m_\chi = 200~\gev$ is noticeably different from  the background model, enabling discriminatory power when energy information is included, and illustrating the capabilities of the ABC analysis technique when energy information is used.

We also consider an alternate scenario, where the dark matter annihilates to $\bar b b$ in Appendix~\ref{app:bbar}. 
For the model assuming the $\bar b b$ annihilation channel and  $m_\chi = 50~\gev$ considered in Appendix~\ref{app:bbar}, the dark matter spectrum is much more similar to the backgrounds at low energies, as seen in Fig.~\ref{fig:pdf_spectrum_models_bbar}.  In this case, we expect less improvement in parameter constraints upon including energy information; this is seen in Fig.~\ref{fig:bbar_posteriors}.

\subsection{Astrophysical background spectral model}
\label{sec:astro_spectra}

As described in \S\ref{sec:astro_counts}, our choice of the amplitude of the background of gamma rays not produced in dark matter annihilation is derived from a combination of the isotropic and galactic Fermi background models that is designed to capture the expected photon statistics.  We set the spectrum of the total background equal to that of the Fermi isotropic background model.  This approach is motivated since we consider only a single background component in our analysis.  Furthermore, the spectra of the isotropic background and the galactic background (over our mask) are very similar, so we expect that ignoring the difference between these two will have a minimal impact on our results.   We illustrate the background spectrum with the red dashed curve in the right panel of Fig.~\ref{fig:pdf_spectrum_models}.  
Over the energy range of interest, this spectrum receives dominant contributions from pion decay and inverse Compton scattering.  As discussed earlier, the convolution of Poisson-distributed sources is trivial, so it is acceptable to simply consider a single energy spectrum for the background which incorporates all approximately Poisson-distributed background source classes.

It is worth noting that, if dark matter annihilation produces a photon signal from subhalos, then it will also produce a Poisson-distributed signal from annihilation within the smooth component of the Milky Way halo, along with a roughly isotropic distribution arising from annihilation outside the Milky Way.  This contribution is part of the data which is fit by the Fermi collaboration to determine the energy-dependent background model; to the extent that this model is a good fit to the data, it incorporates the smooth dark matter component.  This does not present a difficulty.  Dark matter annihilation is expected to provide at most a subleading contribution to this Poisson-distributed background, so it is not surprising that the spectral shape of the background may be very different from that of the dark matter model.   But the subhalos are a source of non-Poisson distributed photons, and knowledge of their energy spectrum will thus provide additional information.

\section{Application of ABC methods}
\label{sec:application}

We now perform a series of analyses of mock data to illustrate that the ABC method works, and that it can result in significantly improved parameter constraints when energy information is included.

\begin{figure*}
    \centering
    \includegraphics[scale=0.5]{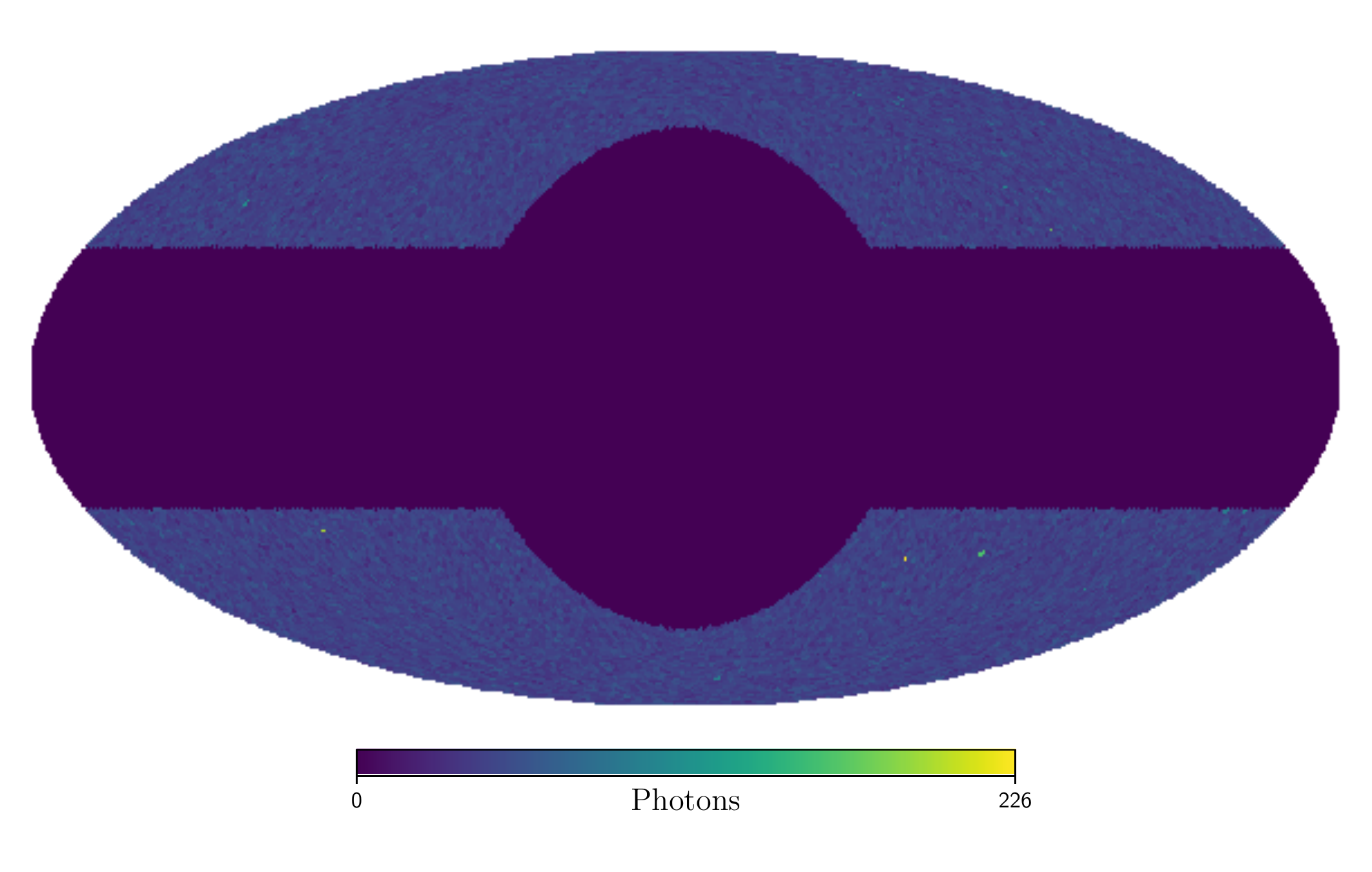}
    \caption{Example mock sky map (summed over all energy bins, and shown in Mollweide projection) including dark matter annihilation and astrophysical background contributions.  The dark band along the center of the image is our mask, which removes parts of the sky close to the galactic plane or towards the galactic center. The dark matter and background models used to generate the mock map correspond to those shown in Fig.~\ref{fig:pdf_spectrum_models}.  The non-Poisson nature of the dark matter signal can be seen via the few very bright pixels. 
    }
    \label{fig:example_map}
\end{figure*}

\begin{figure}
    \centering
    \includegraphics[scale=0.45]{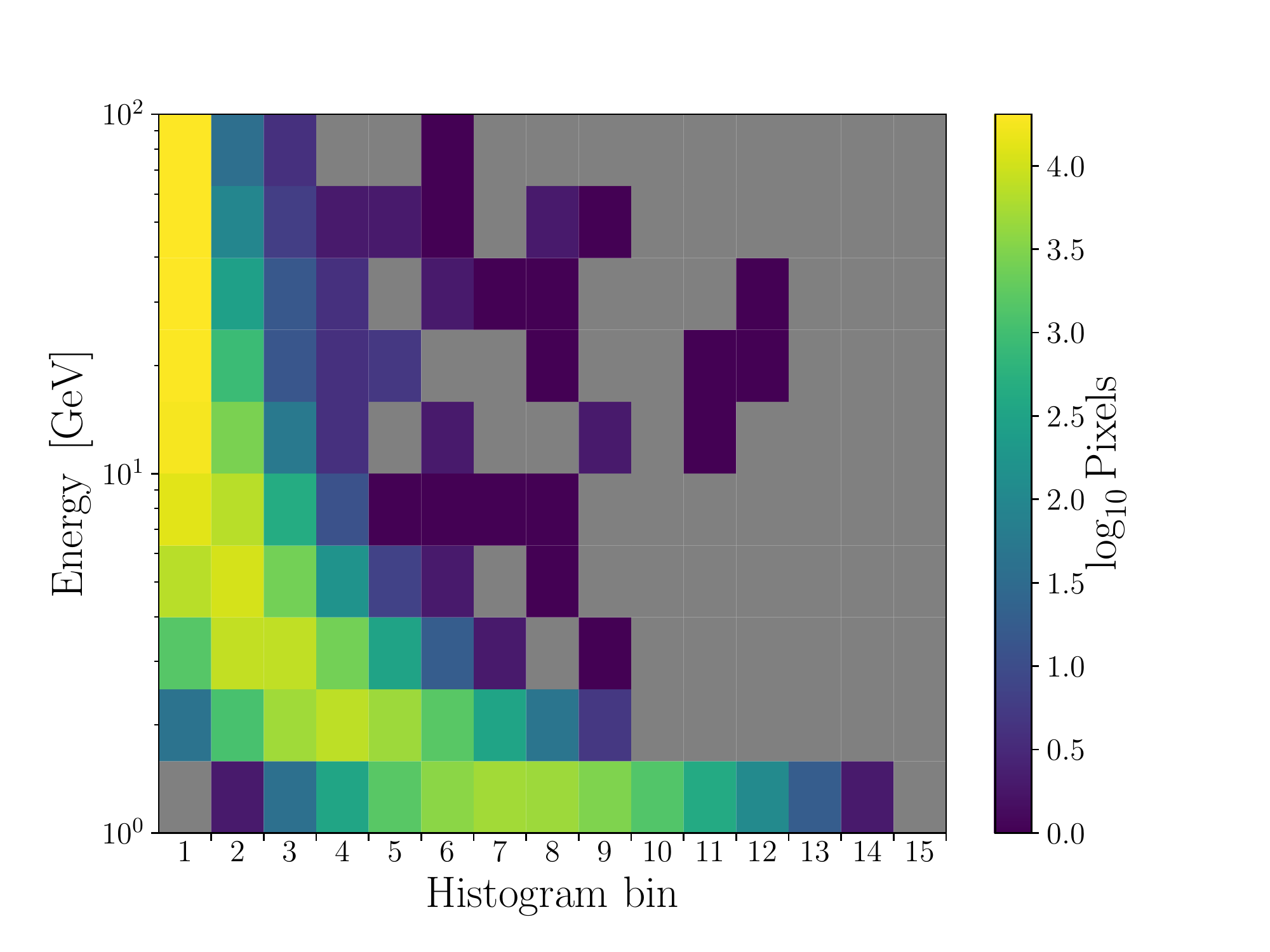}
    \caption{Visual representation of the summary statistic used in our ABC analysis, corresponding to the map shown in Fig.~\ref{fig:example_map}.   We compress the energy-dependent pixelized maps into energy-dependent histograms.  The non-Poisson nature of the counts distribution at high energy is apparent.  As discussed in \S\ref{sec:summary_stat}, the maximum count value used to define the count bins varies with energy.  Grey regions indicate empty bins.  
    }
    \label{fig:summary_stat}
\end{figure}

We generate a mock data set with both astrophysical backgrounds and a dark matter signal  using the models discussed in \S\ref{sec:PDF_models}.  Our model has three parameters, the normalization of the dark matter signal ($A_{DM}$), the dark matter mass ($m_\chi$), and the normalization of the background ($A_{BG}$).  To generate the mock data,  we set $m_\chi = 200~\gev$ (assuming the $\bar \tau \tau$ channel).     We normalize the dark matter signal to $A_{DM} = 200$, which corresponds to $\Phi_{PP} \sim 4 \times 10^{-30} \cm^3 \s^{-1} \gev^{-2}$, roughly the current bound from searches for $s$-wave dark matter annihilation  in dwarf spheroidal galaxies~\citep{Geringer-Sameth:2011wse,Boddy:2018qur,Boddy:2019kuw}.  We assume Fermi-like observations covering the full sky, with an exposure time of 10 years, a detector area of 2000 ${\rm cm}^2$, and a field of view of $1/5$ the sky.  The photon data are generated across 10 energy bins, logarithmically distributed between 1~GeV and 100~GeV.  A sky map of 
the mock data (summed over all photon energies) in the  \texttt{Healpix}\footnote{\url{https://healpix.jpl.nasa.gov/}} pixelization scheme with 49152 pixels (corresponding to $N_{\rm side} = 64$), is shown in Fig.~\ref{fig:example_map}.  

As discussed previously, we use the energy-dependent counts histogram as a summary statistic.  In our baseline analysis, when ignoring energy information, we use 20 counts bins, evenly distributed between zero and 280 counts.  When including energy information, we use 10 energy bins and 15 counts bins, varying the maximum count value with energy (since as noted previously, the photon flux decreases rapidly with energy).  In particular, for the 10 energy bins, we use maximum counts values of $\vec{C}_{\rm max}=(60, 45, 45, 45, 45, 45, 45, 45, 45, 30)$, in order of increasing energy.  Our binning choices for the alternative model considered in Appendix~\ref{app:bbar} have the same dimensionality, but a difference choice of $\vec{C}_{\rm max}$.   The energy-dependent summary statistic corresponding to this map is  is shown in Figure~\ref{fig:summary_stat}.

We run our analysis on a machine with 28 cores, exploiting the fact that the ABC algorithm can be run in parallel.  Posterior estimation is run using five iterations of the ABC PMC algorithm, which generally takes about five hours, and requires around 20,000 draws of mock data.  As noted previously, the curse of dimensionality results in ABC becoming less efficient as the dimensionality of the data vector is increased.  For our energy-independent analysis, the dimension of the summary statistic is 20.  In the energy-dependent analysis, it is $10\times 15$.  Consequently, we expect some degradation of performance for the energy-dependent case, given the much larger dimensionality of the data vector.  Indeed, we confirm this expectation below.  While the sampling efficiency could be improved by reducing the dimensionality of the summary statistics, doing so would result in increased loss of information relative to the full sky maps.

\subsection{No energy information}
\label{sec:abc_noenergy}

We first apply the ABC methodology without including photon energy information, i.e. by using only one energy bin to compute the summary statistic.  In this case, the exact likelihood can be easily computed as described in \S\ref{sec:exact_likelihood}.  The ABC posterior can then be compared to the exact posterior to determine how well the ABC method is performing.  Ideally, we expect the ABC method to recover exactly the same posterior as the exact likelihood approach.

We analyze the mock data set using a two parameter model where the normalization of the dark matter signal and the background are both varied, but the dark 
matter mass is fixed at the true value of 
$m_\chi = 200~\gev$ (and is assumed to annihilate to the $\bar \tau \tau$).  
Since the analysis in this subsection uses only 
one energy bin, the choices of final state channel and dark matter mass are almost completely degenerate with the dark matter signal normalization.\footnote{The degeneracy between e.g. $m_{\chi}$ and $A_{\rm DM}$ is not complete since, for example, sufficiently extreme values of $m_{\chi}$ will lead to no photons within the acceptance range of the detector.  In that case, no value of $A_{\rm DM}$ can compensate for the zeroed signal.}

\begin{figure}
    \centering
    \includegraphics[scale=0.5]{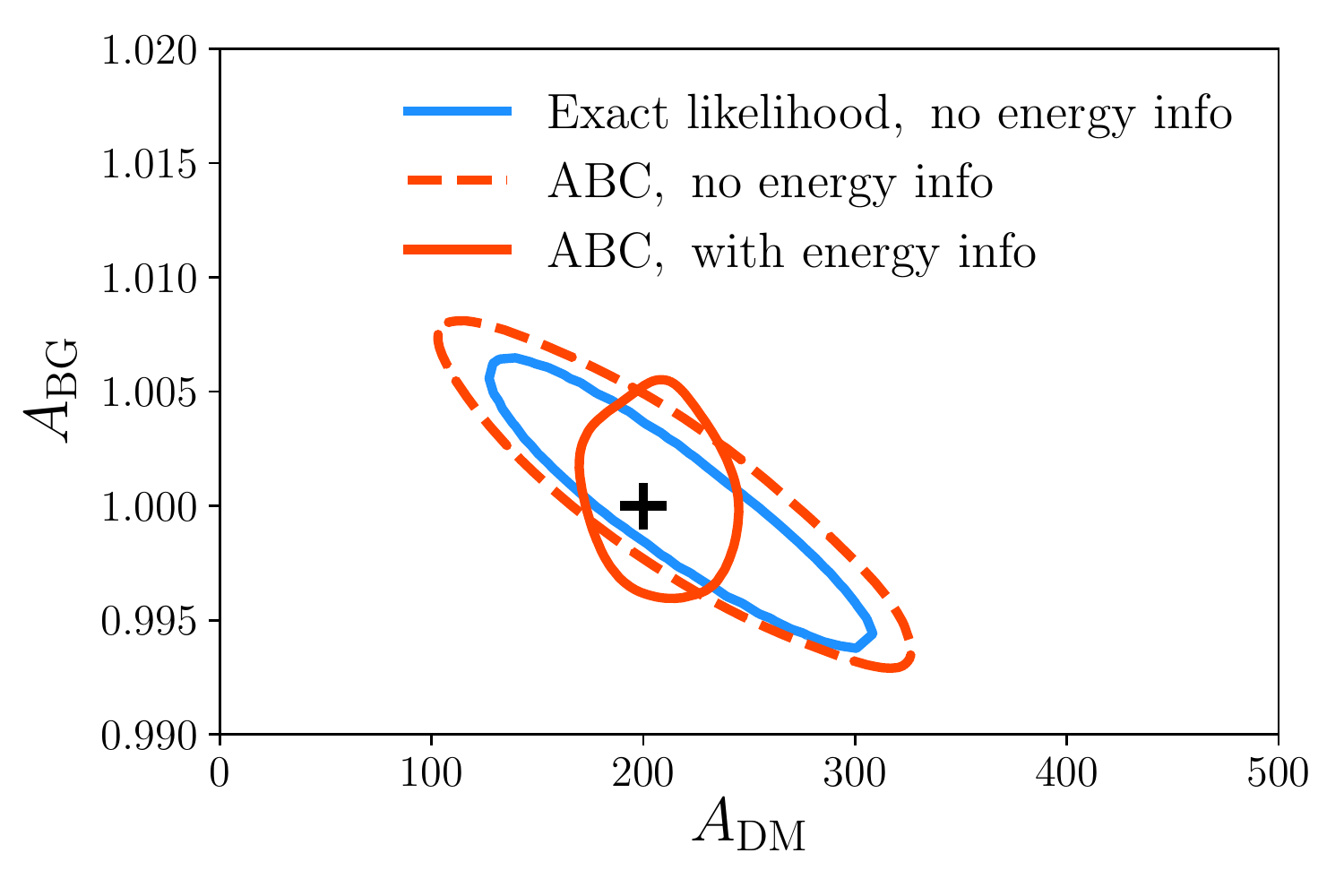}
    \caption{Posteriors (95\% credible intervals) on the dark matter normalization, $A_{\rm DM}$, and the background normalization, $A_{\rm BG}$, recovered using an exact likelihood and using ABC.  Black crosshair indicates the true model parameters.  The exact likelihood (blue solid curve) can only be feasibly computed when energy information is not included.  In this case, the ABC analysis (red dashed curve) recovers a posterior close to the exact likelihood.  When energy information is included --- and the computation of the exact likelihood is no longer feasible --- the ABC method recovers tighter constraints on the model parameters. 
    }
    \label{fig:posterior_comparison}
\end{figure}

The posterior credible intervals (containing 95\% of the posterior mass) that result from analyzing this data using ABC (dashed red) and 
the exact likelihood (blue) are shown in Fig.~\ref{fig:posterior_comparison}.  We see that the exact posterior recovers the input parameter values to within the uncertainties, as expected.  There is also a degeneracy between the dark matter and background amplitudes: increasing the dark matter amplitude must be compensated by a decrease in the background amplitude to preserve the total photon counts.  

Fig.~\ref{fig:posterior_comparison} shows that the posterior from ABC provides a close match to the exact posterior.  It thus appears that the ABC method is working as intended.  The posterior from ABC may be slightly larger than that from the exact likelihood for two reasons.  First, we only expect ABC to recover the exact posterior in the limit that the distance threshold, $\epsilon$, goes to zero, which would require an infinite number of simulated data sets.  Second, the exact likelihood operates on the full map data.  Our implementation of ABC, however, operates on the histogram summary statistic, which may result in some information loss, as discussed in \S\ref{sec:summary_stat}.

\begin{figure}
    \centering
    \includegraphics[scale=0.5]{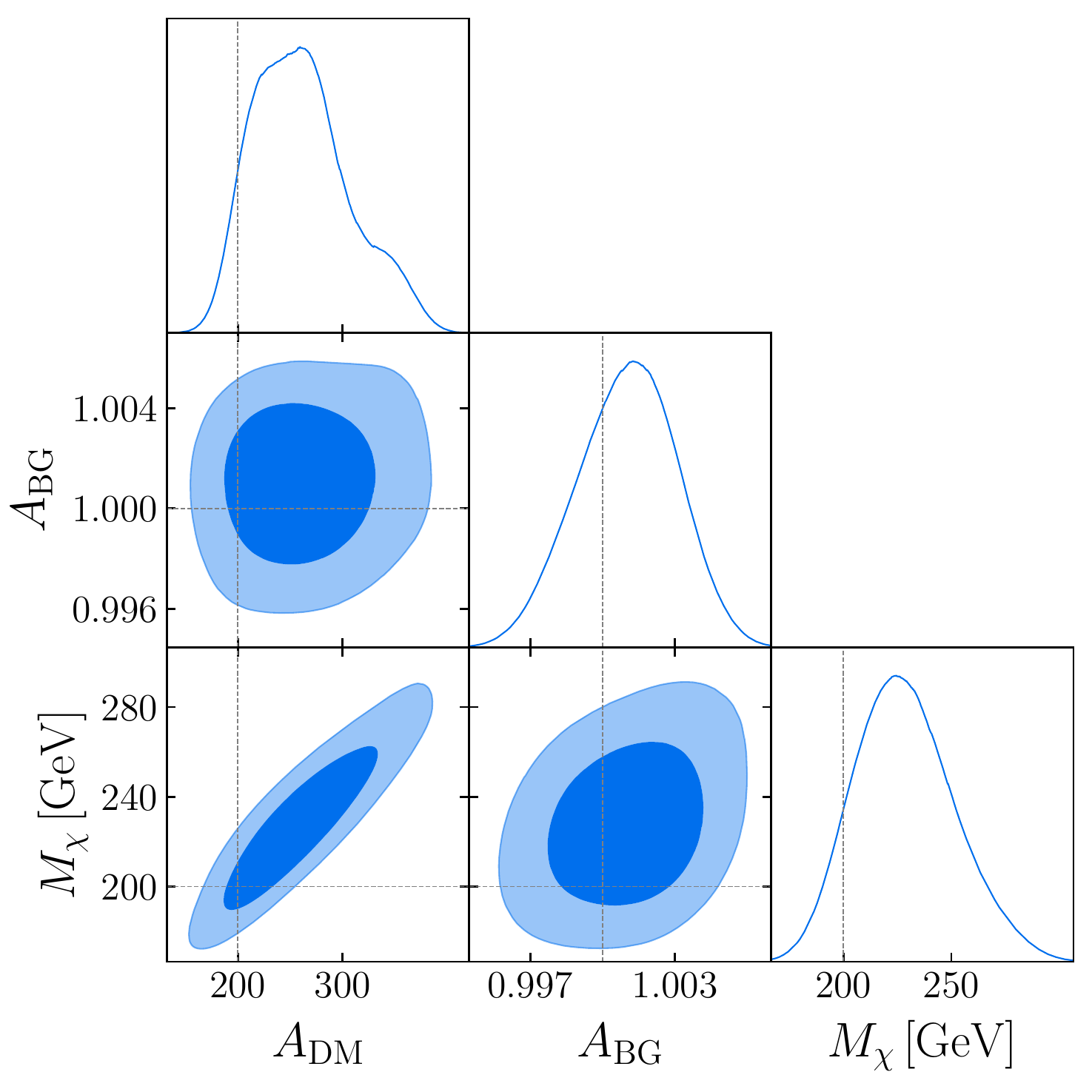}
    \caption{Recovered posterior (68\% and 95\% credible intervals) using ABC when the dark matter mass ($m_{\chi}$) is varied in addition to the normalization parameters $A_{\rm DM}$ and $A_{\rm BG}$.  We include energy information in this analysis.  Grey lines indicate the input parameter values, which are recovered to within the uncertainties.  Without energy information, the dark matter mass and amplitude would be highly degenerate.  While some degeneracy remains here, we break the degeneracy enough to get a constraints on both $A_{\rm DM}$ and $m_{\chi}$. 
    }
    \label{fig:varying_mass}
\end{figure}

\subsection{Improving constraints with energy information and ABC}
\label{sec:abc_withenergy}

We now extend the previous analyses by including energy information.  In this case, the computation of the exact likelihood is intractable, as discussed in \S\ref{sec:exact_likelihood}.  However, the application of the ABC analysis is straightforward.  We begin 
by reanalyzing the same mock data set described in the previous section,  but now using the energy-dependent histogram as our summary statistic.  Again, we keep the true dark matter mass and annihilation 
channel fixed, while varying the normalization of the dark matter signal and the background.  The 
95\% CL posterior obtained from this procedure is shown in Fig.~\ref{fig:posterior_comparison} (solid 
red curve).  We see that including energy information significantly improves our ability to recover the dark matter parameters relative to the case without energy information.  This is one of the main results of our work: using the likelihood free method of ABC, we can easily include energy information in an analysis of the photon count distribution, thereby significantly improving parameter constraints.  
Note that the improvement in constraints lies along the axis in parameter space for which there was a 
degeneracy in the analysis with no energy-information.  Along this direction, the number of photons largely 
remained fixed, but was redistributed between the source classes.  It makes sense that including the energy 
information will help break this degeneracy.  There is no improvement along the orthogonal axis, since this axis 
largely parameterizes the total number of photons, and the energy spectrum provides little additional information.  Given the fact that the dimensionality of the energy-dependent histogram is significantly larger than that of the energy-independent histogram, it is not surprising to see that energy-dependent analysis yields slightly weaker constraints along the orthogonal direction than the energy-independent analysis, simply due to the difference in sampling efficiency.

Since we now include energy information, the choice of dark matter mass and annihilation channel are no longer degenerate with the signal normalization.  To address this, we also consider a three-parameter family of models in which the dark matter mass, as well as the 
dark matter and background normalizations, are allowed to vary, although the 
true annihilation channel is held fixed.  The 95\% CL posteriors on these parameters are presented in Fig.~\ref{fig:varying_mass}.  We see that the data are able to constrain not only the amplitudes of the signal and background, but also the dark matter particle mass.  In other words, by including energy information we have gained access to qualitatively new information about the dark matter.

In Appendix~\ref{app:bbar}, we repeat the analysis described above, but set the dark matter annihilation channel to be $\bar b b$, with the true dark matter mass being chosen as $m_\chi = 50~\gev$.  
This analysis illustrates the improvement that can be obtained by using energy information for the 
case in which the energy spectra of the signal and background are similar.    Not surprisingly, the improvement in parameter constraints is less in this case.

\section{Discussion}
\label{sec:discussion}

We have presented an application of approximate Bayesian computation (ABC) in a mock analysis of the diffuse gamma-ray background (DGRB), demonstrating that it provides significant advantages over traditional likelihood-based approaches.  Using ABC, we are able to recover the exact posterior on our model parameters when energy information from the data is discarded.  When energy information is included, the calculation of the exact likelihood (and thus the posterior) becomes intractable.  However, this limitation does not impact ABC: we are able to recover tight parameter constraints by applying ABC to the energy-dependent photon counts data.  

We have focused on extracting information about dark matter from the DGRB.  Of course, the methods considered here are quite general, and could easily be applied to analyses beyond dark matter annihilation.  In particular, it would be interesting to extend this analysis to include sources such as blazars and star forming galaxies to perform a ``complete'' analysis of the DGRB.  Likelihood-free methods are also applicable to non-istrotropic sources, such as the galactic center gamma-ray excess \citep{Michra-Sharma:2021}. 

Our analysis has ignored several  complications that may impact real data.  For instance, in Fermi data, the exposure varies with position on the sky and with energy, while we have assumed constant exposure.  We have also ignored the impact of uncertainty in the photon directions, which could introduce correlations between nearby pixels.  In Fermi data, the containment angle (i.e. precision with which the photon direction can be determined) also varies with energy.  Our treatment of non-dark matter sources of gamma-ray emission is also highly simplistic, because it assumes isotropy and purely Poisson distributed photon counts.  

We have ignored additional complications related to the specification of the dark matter model.  
For example, we have focused on a class of models in which the energy spectrum is a function of 
a single parameter: the dark matter mass.  More generally, the spectrum could be a function of 
multiple parameters, as would be the case if there are multiple relevant final states, with relative 
branching fractions being parameters determined by the details of the particle physics model.  
Additionally, we have assumed that the parameters of the subhalo mass function can be well-determined 
from numerical simulations.

Fortunately, a significant advantage of ABC is that  observational complications such as varying exposures and correlated pixels can fairly trivially be incorporated into an analysis.  One simply modifies the generation of mock data to include these effects, and repeats the analysis.  
Similarly, one can also  incorporate any additional parameters affecting dark matter microphysics or astrophysics.
For an exact likelihood analysis, on the other hand, including effects such as correlation between pixels is highly non-trivial.

\section*{Acknowledgements}

We thank Scott Dodelson and Aleczander Paul for useful conversations related to this work.   The work of JK is supported in part by DOE grant DE-SC0010504. The technical support and advanced computing resources from University of Hawai‘i Information Technology Services – Cyberinfrastructure, funded in part by the National Science Foundation MRI award \#1920304, are gratefully acknowledged.

\section*{Data availability}

The data used to generate the figures in this work are available upon request.

\bibliographystyle{mnras}
\bibliography{thebib}

\begin{thebibliography}{}
\makeatletter
\relax
\def\mn@urlcharsother{\let\do\@makeother \do\$\do\&\do\#\do\^\do\_\do\%\do\~}
\def\mn@doi{\begingroup\mn@urlcharsother \@ifnextchar [ {\mn@doi@}
  {\mn@doi@[]}}
\def\mn@doi@[#1]#2{\def\@tempa{#1}\ifx\@tempa\@empty \href
  {http://dx.doi.org/#2} {doi:#2}\else \href {http://dx.doi.org/#2} {#1}\fi
  \endgroup}
\def\mn@eprint#1#2{\mn@eprint@#1:#2::\@nil}
\def\mn@eprint@arXiv#1{\href {http://arxiv.org/abs/#1} {{\tt arXiv:#1}}}
\def\mn@eprint@dblp#1{\href {http://dblp.uni-trier.de/rec/bibtex/#1.xml}
  {dblp:#1}}
\def\mn@eprint@#1:#2:#3:#4\@nil{\def\@tempa {#1}\def\@tempb {#2}\def\@tempc
  {#3}\ifx \@tempc \@empty \let \@tempc \@tempb \let \@tempb \@tempa \fi \ifx
  \@tempb \@empty \def\@tempb {arXiv}\fi \@ifundefined
  {mn@eprint@\@tempb}{\@tempb:\@tempc}{\expandafter \expandafter \csname
  mn@eprint@\@tempb\endcsname \expandafter{\@tempc}}}

\bibitem[\protect\citeauthoryear{{Baxter}, {Dodelson}, {Koushiappas}  \&
  {Strigari}}{{Baxter} et~al.}{2010}]{Baxter:2010}
{Baxter} E.~J.,  {Dodelson} S.,  {Koushiappas} S.~M.,   {Strigari} L.~E.,
  2010, \mn@doi [\prd] {10.1103/PhysRevD.82.123511}, \href
  {https://ui.adsabs.harvard.edu/abs/2010PhRvD..82l3511B} {82, 123511}

\bibitem[\protect\citeauthoryear{{Baxter}, {Kumar}, {Paul}  \&
  {Runburg}}{{Baxter} et~al.}{2022}]{Baxter:2022}
{Baxter} E.~J.,  {Kumar} J.,  {Paul} A.~D.,   {Runburg} J.,  2022, arXiv
  e-prints, \href {https://ui.adsabs.harvard.edu/abs/2022arXiv220502386B} {p.
  arXiv:2205.02386}

\bibitem[\protect\citeauthoryear{{Beaumont}, {Cornuet}, {Marin}  \&
  {Robert}}{{Beaumont} et~al.}{2008}]{Beaumont:2008}
{Beaumont} M.~A.,  {Cornuet} J.-M.,  {Marin} J.-M.,   {Robert} C.~P.,  2008,
  arXiv e-prints, \href {https://ui.adsabs.harvard.edu/abs/2008arXiv0805.2256B}
  {p. arXiv:0805.2256}

\bibitem[\protect\citeauthoryear{Boddy, Kumar, Marfatia  \& Sandick}{Boddy
  et~al.}{2018}]{Boddy:2018qur}
Boddy K.,  Kumar J.,  Marfatia D.,   Sandick P.,  2018, \mn@doi [Phys. Rev. D]
  {10.1103/PhysRevD.97.095031}, 97, 095031

\bibitem[\protect\citeauthoryear{Boddy, Hill, Kumar, Sandick  \& Shams
  Es~Haghi}{Boddy et~al.}{2021}]{Boddy:2019kuw}
Boddy K.~K.,  Hill S.,  Kumar J.,  Sandick P.,   Shams Es~Haghi B.,  2021,
  \mn@doi [Comput. Phys. Commun.] {10.1016/j.cpc.2020.107815}, 261, 107815

\bibitem[\protect\citeauthoryear{Cirelli et~al.,}{Cirelli
  et~al.}{2011}]{Cirelli:2010xx}
Cirelli M.,  et~al., 2011, \mn@doi [JCAP] {10.1088/1475-7516/2012/10/E01}, 03,
  051

\bibitem[\protect\citeauthoryear{Cranmer, Brehmer  \& Louppe}{Cranmer
  et~al.}{2020}]{Cranmer:2020}
Cranmer K.,  Brehmer J.,   Louppe G.,  2020, \mn@doi [Proceedings of the
  National Academy of Sciences] {10.1073/pnas.1912789117}, 117, 30055

\bibitem[\protect\citeauthoryear{{Fermi LAT Collaboration}}{{Fermi LAT
  Collaboration}}{2015}]{Fermi:2015}
{Fermi LAT Collaboration} 2015, \mn@doi [\jcap]
  {10.1088/1475-7516/2015/09/008}, \href
  {https://ui.adsabs.harvard.edu/abs/2015JCAP...09..008F} {2015, 008}

\bibitem[\protect\citeauthoryear{{Fornasa} \& {S{\'a}nchez-Conde}}{{Fornasa} \&
  {S{\'a}nchez-Conde}}{2015}]{Fornasa:2015}
{Fornasa} M.,  {S{\'a}nchez-Conde} M.~A.,  2015, \mn@doi [\physrep]
  {10.1016/j.physrep.2015.09.002}, \href
  {https://ui.adsabs.harvard.edu/abs/2015PhR...598....1F} {598, 1}

\bibitem[\protect\citeauthoryear{Geringer-Sameth \&
  Koushiappas}{Geringer-Sameth \& Koushiappas}{2011}]{Geringer-Sameth:2011wse}
Geringer-Sameth A.,  Koushiappas S.~M.,  2011, \mn@doi [Phys. Rev. Lett.]
  {10.1103/PhysRevLett.107.241303}, 107, 241303

\bibitem[\protect\citeauthoryear{{Harding} \& {Abazajian}}{{Harding} \&
  {Abazajian}}{2012}]{Harding:2012}
{Harding} J.~P.,  {Abazajian} K.~N.,  2012, \mn@doi [\jcap]
  {10.1088/1475-7516/2012/11/026}, \href
  {https://ui.adsabs.harvard.edu/abs/2012JCAP...11..026H} {2012, 026}

\bibitem[\protect\citeauthoryear{{Hasinger}, {Burg}, {Giacconi}, {Hartner},
  {Schmidt}, {Trumper}  \& {Zamorani}}{{Hasinger} et~al.}{1993}]{Hasinger:1993}
{Hasinger} G.,  {Burg} R.,  {Giacconi} R.,  {Hartner} G.,  {Schmidt} M.,
  {Trumper} J.,   {Zamorani} G.,  1993, \aap, \href
  {https://ui.adsabs.harvard.edu/abs/1993A&A...275....1H} {275, 1}

\bibitem[\protect\citeauthoryear{{Jimenez Rezende} \& {Mohamed}}{{Jimenez
  Rezende} \& {Mohamed}}{2015}]{Rezende:2015}
{Jimenez Rezende} D.,  {Mohamed} S.,  2015, arXiv e-prints, \href
  {https://ui.adsabs.harvard.edu/abs/2015arXiv150505770J} {p. arXiv:1505.05770}

\bibitem[\protect\citeauthoryear{{Koushiappas}, {Zentner}  \&
  {Kravtsov}}{{Koushiappas} et~al.}{2010}]{Koushiappas:2010}
{Koushiappas} S.~M.,  {Zentner} A.~R.,   {Kravtsov} A.~V.,  2010, \mn@doi
  [\prd] {10.1103/PhysRevD.82.083504}, \href
  {https://ui.adsabs.harvard.edu/abs/2010PhRvD..82h3504K} {82, 083504}

\bibitem[\protect\citeauthoryear{Kumar \& Marfatia}{Kumar \&
  Marfatia}{2013}]{Kumar:2013iva}
Kumar J.,  Marfatia D.,  2013, \mn@doi [Phys. Rev. D]
  {10.1103/PhysRevD.88.014035}, 88, 014035

\bibitem[\protect\citeauthoryear{{Lee}, {Ando}  \& {Kamionkowski}}{{Lee}
  et~al.}{2009}]{Lee:2009}
{Lee} S.~K.,  {Ando} S.,   {Kamionkowski} M.,  2009, \mn@doi [\jcap]
  {10.1088/1475-7516/2009/07/007}, \href
  {https://ui.adsabs.harvard.edu/abs/2009JCAP...07..007L} {2009, 007}

\bibitem[\protect\citeauthoryear{Lintusaari et~al.,}{Lintusaari
  et~al.}{2018}]{ELFI}
Lintusaari J.,  et~al., 2018, Journal of Machine Learning Research, 19, 1

\bibitem[\protect\citeauthoryear{{Madau}, {Diemand}  \& {Kuhlen}}{{Madau}
  et~al.}{2008}]{Madau:2008}
{Madau} P.,  {Diemand} J.,   {Kuhlen} M.,  2008, \mn@doi [\apj]
  {10.1086/587545}, \href
  {https://ui.adsabs.harvard.edu/abs/2008ApJ...679.1260M} {679, 1260}

\bibitem[\protect\citeauthoryear{{Malyshev} \& {Hogg}}{{Malyshev} \&
  {Hogg}}{2011}]{Malyshev:2011}
{Malyshev} D.,  {Hogg} D.~W.,  2011, \mn@doi [\apj]
  {10.1088/0004-637X/738/2/181}, \href
  {https://ui.adsabs.harvard.edu/abs/2011ApJ...738..181M} {738, 181}

\bibitem[\protect\citeauthoryear{{Martinez}, {Bullock}, {Kaplinghat},
  {Strigari}  \& {Trotta}}{{Martinez} et~al.}{2009}]{Martinez:2009}
{Martinez} G.~D.,  {Bullock} J.~S.,  {Kaplinghat} M.,  {Strigari} L.~E.,
  {Trotta} R.,  2009, \mn@doi [\jcap] {10.1088/1475-7516/2009/06/014}, \href
  {https://ui.adsabs.harvard.edu/abs/2009JCAP...06..014M} {2009, 014}

\bibitem[\protect\citeauthoryear{{Mishra-Sharma} \& {Cranmer}}{{Mishra-Sharma}
  \& {Cranmer}}{2021}]{Michra-Sharma:2021}
{Mishra-Sharma} S.,  {Cranmer} K.,  2021, arXiv e-prints, \href
  {https://ui.adsabs.harvard.edu/abs/2021arXiv211006931M} {p. arXiv:2110.06931}

\bibitem[\protect\citeauthoryear{{Papamakarios}, {Nalisnick}, {Jimenez
  Rezende}, {Mohamed}  \& {Lakshminarayanan}}{{Papamakarios}
  et~al.}{2019}]{Papamakarios:2019}
{Papamakarios} G.,  {Nalisnick} E.,  {Jimenez Rezende} D.,  {Mohamed} S.,
  {Lakshminarayanan} B.,  2019, arXiv e-prints, \href
  {https://ui.adsabs.harvard.edu/abs/2019arXiv191202762P} {p. arXiv:1912.02762}

\bibitem[\protect\citeauthoryear{{Pavlidou} \& {Fields}}{{Pavlidou} \&
  {Fields}}{2002}]{Pavlidou:2002}
{Pavlidou} V.,  {Fields} B.~D.,  2002, \mn@doi [\apjl] {10.1086/342670}, \href
  {https://ui.adsabs.harvard.edu/abs/2002ApJ...575L...5P} {575, L5}

\bibitem[\protect\citeauthoryear{Pele \& Werman}{Pele \&
  Werman}{2010}]{Pele:2010}
Pele O.,  Werman M.,  2010, in Daniilidis K.,  Maragos P.,   Paragios N.,  eds,
  Computer Vision -- ECCV 2010. Springer Berlin Heidelberg, Berlin, Heidelberg,
  pp 749--762

\bibitem[\protect\citeauthoryear{{Roth}, {Krumholz}, {Crocker}  \&
  {Celli}}{{Roth} et~al.}{2021}]{Roth:2021}
{Roth} M.~A.,  {Krumholz} M.~R.,  {Crocker} R.~M.,   {Celli} S.,  2021, \mn@doi
  [\nat] {10.1038/s41586-021-03802-x}, \href
  {https://ui.adsabs.harvard.edu/abs/2021Natur.597..341R} {597, 341}

\bibitem[\protect\citeauthoryear{Rubin}{Rubin}{1984}]{Rubin:1984}
Rubin D.~B.,  1984, \mn@doi [The Annals of Statistics]
  {10.1214/aos/1176346785}, 12, 1151

\bibitem[\protect\citeauthoryear{{Runburg}, {Baxter}  \& {Kumar}}{{Runburg}
  et~al.}{2021}]{Runburg:2021}
{Runburg} J.,  {Baxter} E.~J.,   {Kumar} J.,  2021, arXiv e-prints, \href
  {https://ui.adsabs.harvard.edu/abs/2021arXiv210610399R} {p. arXiv:2106.10399}

\bibitem[\protect\citeauthoryear{{Scheuer}}{{Scheuer}}{1957}]{Scheuer:1957}
{Scheuer} P.~A.~G.,  1957, \mn@doi [Proceedings of the Cambridge Philosophical
  Society] {10.1017/S0305004100032825}, \href
  {https://ui.adsabs.harvard.edu/abs/1957PCPS...53..764S} {53, 764}

\bibitem[\protect\citeauthoryear{{Simola}, {Cisewski-Kehe}, {Gutmann}  \&
  {Corander}}{{Simola} et~al.}{2019}]{Simola:2019}
{Simola} U.,  {Cisewski-Kehe} J.,  {Gutmann} M.~U.,   {Corander} J.,  2019,
  arXiv e-prints, \href {https://ui.adsabs.harvard.edu/abs/2019arXiv190701505S}
  {p. arXiv:1907.01505}

\bibitem[\protect\citeauthoryear{Sisson, Fan  \& Tanaka}{Sisson
  et~al.}{2007}]{Sisson:2007}
Sisson S.~A.,  Fan Y.,   Tanaka M.~M.,  2007, \mn@doi [Proceedings of the
  National Academy of Sciences] {10.1073/pnas.0607208104}, 104, 1760

\bibitem[\protect\citeauthoryear{Somalwar, Chang, Mishra-Sharma  \&
  Lisanti}{Somalwar et~al.}{2021}]{Somalwar:2020awt}
Somalwar J.~J.,  Chang L.~J.,  Mishra-Sharma S.,   Lisanti M.,  2021, \mn@doi
  [Astrophys. J.] {10.3847/1538-4357/abc87d}, 906, 57

\bibitem[\protect\citeauthoryear{{Stecker}, {Salamon}  \& {Malkan}}{{Stecker}
  et~al.}{1993}]{Stecker:1993}
{Stecker} F.~W.,  {Salamon} M.~H.,   {Malkan} M.~A.,  1993, \mn@doi [\apjl]
  {10.1086/186882}, \href
  {https://ui.adsabs.harvard.edu/abs/1993ApJ...410L..71S} {410, L71}

\bibitem[\protect\citeauthoryear{{Xia}, {Cuoco}, {Branchini}, {Fornasa}  \&
  {Viel}}{{Xia} et~al.}{2011}]{Xia:2011}
{Xia} J.-Q.,  {Cuoco} A.,  {Branchini} E.,  {Fornasa} M.,   {Viel} M.,  2011,
  \mn@doi [\mnras] {10.1111/j.1365-2966.2011.19200.x}, \href
  {https://ui.adsabs.harvard.edu/abs/2011MNRAS.416.2247X} {416, 2247}

\bibitem[\protect\citeauthoryear{{Zhao}, {Hooper}, {Angus}, {Taylor}  \&
  {Silk}}{{Zhao} et~al.}{2007}]{Zhao:2007}
{Zhao} H.,  {Hooper} D.,  {Angus} G.~W.,  {Taylor} J.~E.,   {Silk} J.,  2007,
  \mn@doi [\apj] {10.1086/509649}, \href
  {https://ui.adsabs.harvard.edu/abs/2007ApJ...654..697Z} {654, 697}

\makeatother
\end{thebibliography}

\appendix

\section{Alternate annihilation model}
\label{app:bbar}

\begin{figure*}
    \centering
    \includegraphics[scale=0.48]{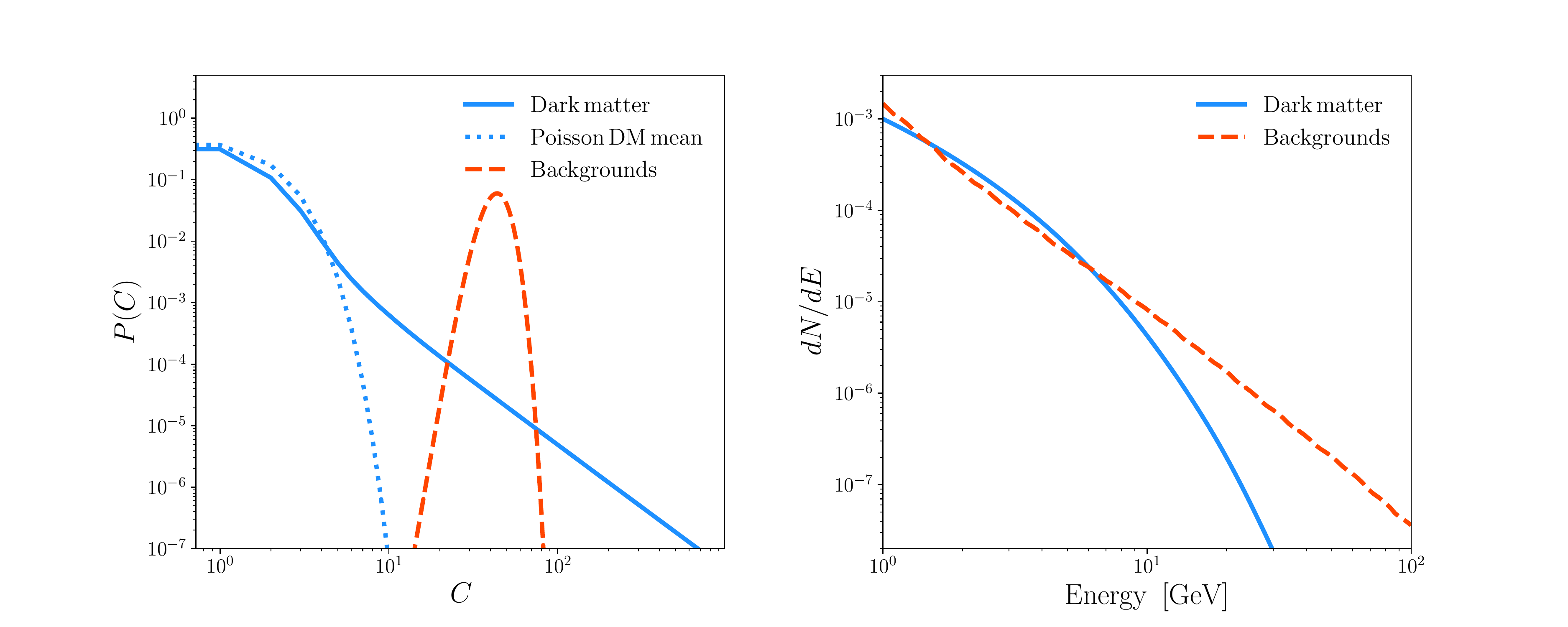}
    \caption{Illustration of the photon counts distributions (left) and spectral models (right) for dark matter with $m_{\chi} = 50\,{\rm GeV}$ that annihilates to $b\bar{b}$ (blue solid) and backgrounds (red dashed).  The blue dotted curve indicates a Poisson counts distribution model with the same mean as the dark matter signal.}
    \label{fig:pdf_spectrum_models_bbar}
\end{figure*}

In the main text, we considered dark matter with $m_{\chi} = 200\,{\rm GeV}$ that annihilates to $\tau\bar{\tau}$.  In that case, the photon spectrum from dark matter annihilation differs significantly from that of the astrophysical backgrounds.  This meant that adding energy information to the analysis resulted in significantly improved parameter constraints.  Here, we consider an alternate model that has $m_{\chi} = 50,{\rm GeV}$, and where the dark matter annihilates to $b\bar{b}$.  The photon count PDFs and spectra for this model are shown in Fig.~\ref{fig:pdf_spectrum_models_bbar}.  We see that for low energies ($E \lesssim 10\,{\rm GeV}$), the annihilation and background spectra are very similar.  Since most photons in this model are produced at low energy, this means that including energy information is not expected to result in significantly improved parameter constraints.  For the $b\bar{b}$ analysis, our energy-independent histogram uses 60 bins between 0 and 720 photons; our energy-dependent histogram uses 10 energy bins and 15 counts bins, with maximum counts given by $\vec{C}_{\rm max} = (315, 180, 105, 60, 45, 30, 15, 15, 15, 15 )$, in order of increasing energy.

\begin{figure}
    \centering
    \includegraphics[scale = 0.5]{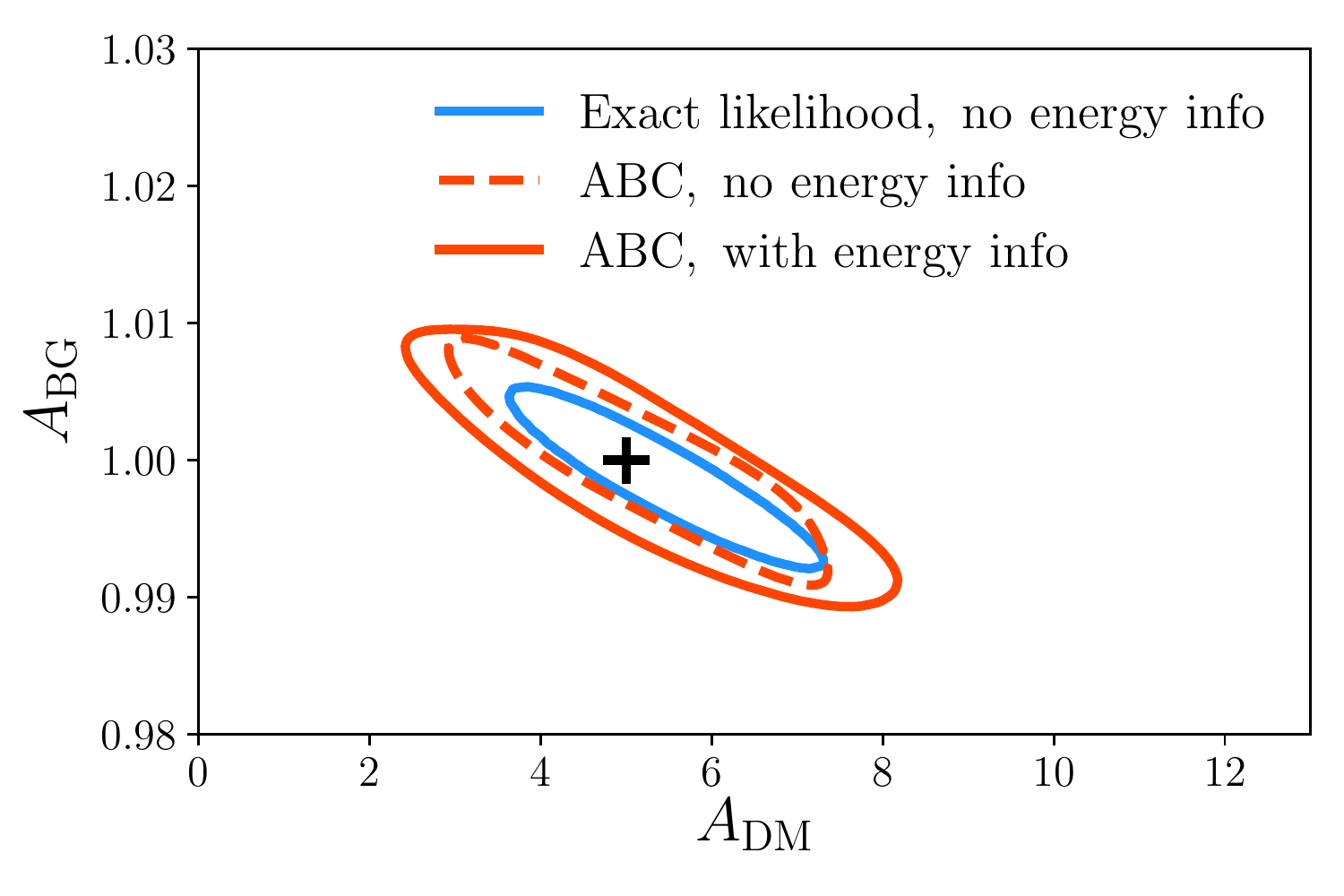}
    \caption{Posteriors (95\% credible intervals) obtained when the true model is dark matter with $m_{\chi} = 50\,{\rm GeV}$ that annihilates to $b\bar{b}$.  In this case, since the spectrum of annihilation radiation is very similar to that of the astrophysical backgrounds for $E \lesssim 10\,{\rm GeV}$ --- the energy range over which most photons are produced --- the improvement upon adding energy information is negligible. }
    \label{fig:bbar_posteriors}
\end{figure}

In Fig.~\ref{fig:bbar_posteriors} we show the posteriors that result from analyzing the mock data with (1) the exact likelihood and no energy information (blue solid), (2) ABC and no energy information (red dashed), (3) ABC and energy information (red solid).  
Note that the exact posterior without energy information is the same as for the $\bar \tau \tau$ case (Fig.~\ref{fig:posterior_comparison}), up to a rescaling of $A_{DM}$.  This is expected, as in the absence of energy information, $A_{DM}$ is degenerate with the choice of channel.  As expected, ABC recovers close to the exact posterior in the absence of energy information.  When energy information is included, there is no significant improvement since the signal and background spectra are close to degenerate.  Indeed, the posterior upon including energy information is actually somewhat broader than the posterior that ignores energy information; this is likely due to less optimal sampling when the dimensionality of the summary statistic is high, which is the case for the analysis that includes energy information.

\bsp	
\label{lastpage}
\end{document}